# Co-operative Phenomena as a Physical Paradigm for Relativity, Gravitation and Quantum Mechanics

Vincent Buonomano[*]

We take co-operative phenomena as a common physical conceptual base to speculate on the existence of a medium and the properties that it must have to physically understand some of the problems in Special Relativity, gravitation and quantum mechanics.

## Introduction

We would like to speculate on the existence of a preferred physical universal reference frame, that is, a medium, and its structure. The basic purport of this work is that the area of co-operative phenomena (Refs. 1-2) may furnish a common physical conceptual basis to better understand gravitation, Special Relativity and quantum mechanics. The medium, the patterns and the equilibrium states that are formed in it are taken as the fundamental concepts to understanding the physical world. Some of these speculations began in earlier works (Refs. 3 and 4).

Before continuing with this task we think it appropriate to recall the following. The word medium is frequently taken as a synonym for the word ether. The ether was an attempt to try and justify a medium as a mechanical fluid to support the transverse vibrations of electromagnetic waves in a manner completely analogous to how waves are justified in a fluid. That is, the attempt to justify an ether was in terms of what was then considered be the more basic theory or paradigm, basically that of Newtonian mechanics. This attempt occupied some of the best scientists of the second part of the last century. It was a dismal failure, of course. This without doubt gave a 'bad name' and an unsophisticated sense to any research involving a medium.

Today it is clear that the concepts from stochastic processes and co-operative phenomena give us much more general structures from which to try to justify and physically understand a medium than does mechanics. There are profound conceptual simplifications in having a medium. Our own opinion is that our discarding the concept of a medium along with the failed classical fluid (the ether) has forced us to give up some essential absolute concepts and reference points. This constrains us to use more complicated conceptual constructs analogous to Ptolemy's geocentric astronomical system compared with the Copernican system.

---

[*] Instituto de Matemática (Retired), Universidade Estadual de Campinas, Campinas, São Paulo, Brazil, 13081-970, e-mail: vincent@correionet.com.br

In the sections below we describe the rough physical properties that we feel one needs to confront some of the difficulties in modern physics. The properties that we want to impose on this medium, without doubt, relate to our own view of these problems. They are speculations, nothing more than physical ideas that may find justification.

## Gravitation

We assume that the medium consists of unknown particles in stochastic motion, and we simply will call them medium particles. All other particles like electrons, neutrons, atoms, *etc.*, will be called normal particles[*]. The number of medium particles is taken to be extremely large and that they permeate all other matter. The medium particles are in constant collision with each other in some sort of stochastic equilibrium, described by their normalized density, $p(x,v) = p(x,v,t)$, in position, $x$, and velocity, $v$. Let $p(x) = \int p(x,v)dv$ and $p(v) = \int p(x,v)dx$. The reference system in which the average stochastic velocity of the medium particles is zero is taken to define a preferred or absolute reference system, S.

In constrast to Le Sage[†] and others we assume that concepts like mass and momentum do not apply to these medium particles. The number and frequency of collisions of normal particles with these medium particles will define the concepts of mass and momentum but only for normal particles. We are necessarily vague about what happens when two medium particles collide, other than that the collisions are impenetrable and that their before-and-after velocities must only be conserved on average.

### Gravitational Force

Let m be a small test mass at rest in the medium at some point $x$. That is, on average, its velocity is zero in any direction in relationship to the preferred reference system. If m is far from any mass distribution, then $p(x)$ is taken as constant independent of $x$, and $p(v)$ is taken to be some fixed unknown distribution independent of $x$.

---

[*] What is a normal particle? Is it made up of medium particles? Is it, itself, a very stable pattern in the medium, something like in the Game of Life? Although this question is fundamental, it is largely irrelevant in this work.

[†] The work of Le Sage will be amply documented in other works in this volume honouring him.

Let us then place a large, spherical mass $M$ at some distance from $m$, also at rest. Then $m$ is no longer in an isotropic medium because of the blocking or shadowing effect that $M$ has, completely analogous to any Le Sage theory. That is, $p(x,v)$ is altered by the mass $M$ and is symmetric about $M$. Our test mass, m, is then no longer in an isotropic medium. It falls toward the mass $M$ because it suffers more collisions on one side than on the other because of a gradient in position and velocity density of the medium. Gravitational force is taken be the greater probability of experiencing collisions in certain directions than others. If the masses are not at rest then one must imagine that the same logic may be applied because of the physics behind the Lorentz transformations. Any Le Sage theory must also confront the question of non-linearity at some point or other because of the shielding or shadowing effect of the shielding on itself.

## Inertial and Gravitational Mass

It is important here to imagine that if a single normal particle and a single medium particle collide then the laws of mechanics governing collisions do not apply. It makes no sense here to say that the normal particle, being larger, 'pushes' the medium particles more than *vice versa*. Mass and momentum are not intrinsic properties of the medium particles or normal particles. Normal particles acquire these properties because of the large number of collisions with the medium particles and the equilibrium of the medium particles with themselves.

Both gravitation and inertial mass are both taken to be basically a property of the size of the object. Gravitational mass is the disturbance that an object produces in the medium in the manner of Le Sage. Inertial mass is its resistance to acceleration because of the greater number of collisions in the direction of the acceleration.

## Probabilistic Potential Theory

In the 1950's the fundamental mathematical result that potential theory and stochastic processes are isomorphic was formalized (Ref. 5). That is, every theorem in potential theory had a corresponding theorem in stochastic processes and vice versa[*]. In particular, the Newtonian potential corresponds *exactly* to Brownian motion. Letting $V(x)$ be the Newtonian potential of the mass distribution $M(y)$, one expresses this by the formulae

---

[*] There are various mathematical regularity conditions that we are omitting here.

$$V(x) = \frac{1}{2\pi} \int \frac{1}{|x-y|} M(y) dy$$

$$= \iint P(y, 0 \to x, t) M(y) dy dt$$

where $P(y,0 \to x,t)$ is the probability a particle undergoing Brownian motion at $y$ at time 0 is at $x$ at time $t$. The integral in $t$ is from 0 to infinity and must be understood as a limit over larger and larger finite time intervals. $P$ is the probability that a particle in Brownian motion goes from $y$ to $x$ in time $t$,

$$P(y,0 \to x,t) = \frac{1}{(\pi 2 Dt)^{3/2}} e^{-(x-y)^2 / 2Dt},$$

where $D$ is the unknown diffusion constant. In words, within the stochastic paradigm, one would say that the potential $V(x)$ at a point $x$ from a single point mass $M$ at $x = 0$ is the sum of the probabilities over a long time period that particles in Brownian motion starting at $x = 0$ arrive at $x$ weighted by $M$.

Presumably one should relate this $D$ to the diffusion coefficients $D_{xx}(x,v)$ of the stochastic process $p(x,v,t)$ representing our medium in the following manner. Let $P(x,v,t \to x',v',t')$ be the transition probability density of the stochastic medium. Here $x$, $v$, $x'$ and $v'$ are three-dimensional vectors. By definition

$$D_{xx}(x,v) = \mathrm{Lim}_{t' \to t} \frac{1}{t'-t} \int (x'-x)^2 P(x,v,t \to x',v',t') dx' dv'.$$

We then take D as equal to $\int D(x,v) dv$ independent of $x$. Further it is natural to want to try to show that the previous would be consistent with taking $p(x,v,t)$ to satisfy the time independent Fokker-Planck equation and, in particular, detailed balancing. It is also natural here to want to consider energy to be directly related to probability, and that conservation of energy expresses the conservation of probability.

## Special Relativity

It is well known[*] that all the effects of Special Relativity may be derived from any theory which assumes the following: (1) There is a preferred or absolute reference frame, S, and that the velocity of light has the constant value c relative to it; (2) A rod undergoes a real physical contraction given by

$$\Delta l_0 \sqrt{1 - \frac{v^2}{c^2}}$$

as a function of its velocity $v$ relative to S, where $\Delta l_0$ is the initial length in S. (3) A clock's rhythm undergoes a real physical dilation given by

$$\frac{\Delta t_0}{\sqrt{1-\frac{v^2}{c^2}}}$$

as a function of its absolute velocity $v$. This suffices to derive the Lorentz transformation between two arbitrary inertial reference systems and therefore all the effects of Special Relativity. Of course, any such physical theory must justify the medium and why clocks and rods behave like this.

---

[*] For example, see Reference 4.

Here, as in any Lorentzian or preferred reference frame theory, time and space are our classical understanding of these concepts, that is, pre Einstein's theory of Special Relativity. Time and space are independent concepts and not the space-time of Einstein. Of course, it is another thing to give concrete experimental significance to these real contractions. In real experiments space are time are always mixed up. We believe it impossible to distinguish between a preferred reference frame theory and Einstein's within the domain of the effects of the Special Theory of Relativity itself.

### Why do rods contract?

A rod is a very large collection of atoms, which in principle may be described by quantum mechanics. To make our point let us consider our rod to consist of only one atom at rest in the medium. We know from quantum mechanics that it is actually undergoing a complex movement with its average velocity being zero. The extent or size of these movements is characterized partially by its standard deviation, which is determined by its state preparation. We associate the length of m with this standard deviation. We assume that it is this standard deviation that is somehow contracting because of its stochastic movement when it has an absolute velocity $v$. The work of Cane (Refs. 6-7) seems relevant here. She has derived a Lorentzian factor for the change in the standard deviation of a Brownian particle on a one- dimensional lattice with a velocity $v$. That is, the probability to go right is greater than the velocity to left by a certain factor, such that the mean position of the particle moves with the velocity $v$. Her proof is not directly applicable here since the Lorentzian factor does not apply to the initial distribution but to its spreading in time. If we consider a clock to be validly represented by a rod with mirrors at each end in which the number of cycles represents the time, then the time dilation assumption may be justified (if the length contraction may be justified *à la Cane*).

### Quantum Mechanics

The stochastic interpretation of quantum mechanics is well known (Ref. 8). It is mathematically equivalent to quantum mechanics and expresses the Schrödinger equation as a Fokker-Planck equation. It imagines that a particle is undergoing constant stochastic motion. The probabilistic density function, entrance and exit velocities of this stochastic motion are calculable from the wave function, $\Psi(x,t)$, and *vice versa*. The diffusion constant $D = \hbar/m$.

State preparation, the double slit experiment and the non-locality controversy will now be discussed. More details are given in Ref. 9.

**State Preparation**

It is important to say that when you prepare a quantum mechanical state, you are not preparing the state of the particle but the system, that is, the medium *and* the particle. It is invalid to separate the medium here. For example, compare the state preparations of our test particle, m, in two different Gaussian states with, say, the second state having a width ten times the value of the first state. In the first case what you have prepared is a particle along with a local pattern in the medium that accompanies it. This pattern has a certain average size, that of the wave packet of the prepared state. In the second case the particle is in a different equilibrium or order with the system. The pattern, or the local equilibrium, that is traveling with the particle in the medium is 10 times larger then in the previous case. You have prepared the system consisting of the particle and the traveling stable pattern in the medium. The stochastic medium's properties depend in general on the state preparation.

One of the consistent criticisms of the stochastic interpretation of quantum mechanics is that the stochastic process of any test particle, *m*, will be non-Markovian. This simply means the properties of the imagined stochastic medium depend on the state preparation. This criticism is clearly answered here.

**Inertial Reference Systems**

A very old and difficult question is why are inertial systems special? Another way of asking this is, why can a particle undergo rectilinear motion without any energy consumption, but to accelerate it one must spend energy? Within our view we would have to say that the medium supports certain stable equilibrium states with the particle and not others. Here the difference between an inertial and non-inertial state would be the difference between an equilibrium and non-equilibrium state. The inertial velocities are modes of the system. In the case of an inertial particle we know its quantum mechanical velocity distribution must be constant from quantum mechanics. This distribution is somehow stable in the medium.

The velocity of light is considered to be a limiting velocity at which the medium can support certain modes or equilibrium states with normal particles.

**Global Patterns in the Medium**

We want to assume that our medium can form various global periodic patterns or modes, in some currently undefined sense, which are completely stable. The exact pattern in a given region will depend on the physical objects in the region. The relevant objects here, such as slits, beam splitters, mirrors, polarizers, parts of the state preparation, *etc.*, determine the pattern or mode of the medium, *i.e.*, the global state. The possible paths or ways in which the medium can interact with itself must be taken as important when contrasted with classical boundary value problems.

We imagine that the particles of the medium co-operate, that is, they organize themselves into a global pattern depending on the objects (mirrors, slits, *etc.*) in it. The size of the region of a pattern or mode is taken to be at least the size of the coherence volume of all the possible beams in a given first or second order interference experiment. One might want to think of something analogous to the Bernard's rolls[*] existing in shallow, slowly heated water, but not consuming energy. Normal particles are guided by the medium.

**The Double Slit Experiment**

If you make two slits in a screen then after a certain relaxation time (it must be taken to be very fast) the medium will enter a new mode, that is, reorganize itself into a new co-operative state. The exact new state or mode will depend on the size of the slit and the distance between the slits, not on the material of the slit (*e.g.*, paper or lead). If you block one slit or the other then the mode would be different than if both were open. It is more important here to imagine that the slits determine the way the medium can interact with itself to enter a new stable state or mode. A photon (or electron or neutron) passes through one slit only. It 'knows' if the other slit is open or closed from the global mode of the medium. The medium would be in a different mode if only one slit was open. It is being guided by the pattern in the medium.

In the rotor experiment (Ref. 10) one must imagine that the medium enters into a one-arm mode and a two-arm mode consecutively for part of each revolution. The medium must be taken to enter into a mode magnitudes faster than the rotation of the rotor.

**The Correlation Experiments**

Bell's work in 1964 did much to clarify the situation in the foundations of quantum mechanics (Ref. 11). He defined a class of theories, called local realistic theories, which must disagree with quantum mechanics in certain experiments. These experiments measure the correlation of measurements (also called second order interference) made on two particle systems at far sides of an apparatus. The quantum mechanical predictions are non-local but cannot be controlled (*i.e.*, used to make a telephone).

Here one imagines that the states of the measuring apparatus on both sides of an experiment to measure the correlations are completely correlated because of the medium. If we change the angle of a polarizer on one side of the apparatus, then the medium is forced into a new global stable state or mode which includes the other polarizer (mirrors, slits...) on the other side of the apparatus. The whole apparatus must be described by one global state or mode. First and second order interference are explained here in the same manner.

---

[*] For example, see Ref. 2, Page 3.

Of the more than 20 correlation experiments preformed to date which confirm quantum mechanical versus local realistic theoretical predictions only Aspect's experiment (Ref. 12) has attempted to measure some sort of communication between the two sides of a correlation apparatus. A series of others are in progress (Ref. 13). Aspect, with some restrictions, eliminated communication up to the velocity of light. To agree with this experiment one would have to assume that the objects of the medium are magnitudes faster than the velocity of light in order for our complex system, *i.e.*, the medium, to enter into a new equilibrium or ordered global state. This view is experimentally testable in a variation of Franson's experiment (Ref. 9).

## Summary

We have imagined that there exists a stochastic medium whose elementary or medium particles do not carry any momentum in and of themselves (in contrast to Le Sage's particles). The force of gravity caused by a mass, $M$, is taken to result from a variation in the position and velocity density of these medium particles about $M$. This makes it more probable for an object to have a net movement in the direction of $M$. This is why Newton's apple falls here. The exact mathematical relationship between Brownian motion and the Newtonian potential in probabilistic potential theory is taken to give some mathematical credibility to this position. One must assume that the particles have impenetrable collisions with their before-and-after velocities being conserved only on average.

Further it was imagined that what we call the size of an object (a macroparticle) is the standard deviation of its stochastic movement according to laws of stochastic quantum mechanics. Rods contract physically as a function of their velocity relative to the medium because their standard deviation of this movement has this property. The Lorentz transformations may be derived from this and the assumption that the velocity of light is constant relative to S.

In an experimental apparatus to measure first or second order interference effects, we imagine that there exists a stable global pattern that is at least the size of the coherence volume of all the involved beams. If you change the position of a mirror, beam splitter, polarizer, state preparation, *etc*., or block a beam, then a new and different stable global state is entered very quickly. The medium particles must be taken to have superluminal velocity to be consistent with Aspect's experiment. It is experimentally testable as a local realistic theory in a variation of Franson's experiment. It is necessary to understand that the quantum mechanical state preparation prepares both the medium and particle, that is the equilibrium, or ordered state, between them. You cannot separate the medium from the particle here.

The velocity of light is considered to be a limiting velocity at which the medium can support certain modes or equilibrium states with normal particles. An inertial motion of a particle is taken to be one of these modes or equilbrium states.